\documentclass[%
 amsmath,amssymb,
 aps,pre,twocolumn
]{revtex4-2}
\usepackage{graphicx}
\usepackage{physics}
\usepackage{mathtools}
\usepackage{xcolor}
\usepackage{siunitx}

\newcommand{\apdx}[0]{End Matter}
\newcommand{\E}[1]{\mathbb{E} \left[{#1}\right]}
\newcommand{\avg}[1]{\left \langle #1 \right \rangle}
\newcommand{\Eq}[1]{Eq.~\ref{#1}}
\newcommand{\Eqs}[1]{Eqs.~\ref{#1}}
\newcommand{\Fig}[1]{Fig.~\ref{#1}}
\newcommand{\Figs}[1]{Figs.~\ref{#1}}

\newcommand{\avgs}[2]{\left \langle {#1}\cdot\bo{D}^{-1}\cdot{#2} \right \rangle}
\newcommand{\avgg}[1]{\left \langle {#1}\cdot\bo{D}^{-1}\cdot{#1} \right \rangle}

\usepackage{xr}

\usepackage[hidelinks]{hyperref}
\usepackage{cleveref}
\newcommand{\bo}[1]{\mathbf{#1}}
\usepackage{bm}

\begin{document}

\title{Principled model selection for stochastic dynamics}

\author{Andonis Gerardos}
\affiliation{Aix Marseille Univ, CNRS, CINAM, Turing Center for Living Systems, Marseille, France}
\author{Pierre Ronceray}
\email{pierre.ronceray@univ-amu.fr}
\affiliation{Aix Marseille Univ, CNRS, CINAM, Turing Center for Living Systems, Marseille, France}

\begin{abstract}
     Complex dynamical systems, from macromolecules to ecosystems, are often modeled by stochastic differential equations. To learn such models from data, a common approach involves sparse selection among a large function library. However, we show that overfitting arises not just from individual model complexity, but also from the combinatorial growth of possible models. To address this, we introduce Parsimonious Stochastic Inference (PASTIS), a principled method combining likelihood-estimation statistics with extreme value theory to suppress superfluous parameters. PASTIS outperforms existing methods and reliably identifies  minimal models, even with low sampling rates or measurement error. It extends to stochastic partial differential equations, and applies to ecological networks and reaction-diffusion dynamics.
\end{abstract}

\maketitle

\label{sec:intro}

Data-driven approaches to physical modeling, which seek to derive governing equations directly from experimental data, have been rapidly advancing~\cite{northReviewDataDrivenDiscovery2023}. They are of particular interest in the context of dynamical systems, where data are trajectories, whose temporal evolution is modeled by differential equations. We can distinguish different levels of ambition for such data-driven methods, ranging from the estimation of parameters of a known equation~\cite{barFittingPartialDifferential1999,ramsayParameterEstimationDifferential2007}, to the discovery of a minimal model among a large class of possible ones. For deterministic systems governed by ordinary or partial differential equations, the advent of symbolic regression~\cite{schmidtDistillingFreeFormNatural2009,cranmerDiscoveringSymbolicModels2020} and Sparse Identification of Nonlinear Dynamics (SINDy)~\cite{bruntonDiscoveringGoverningEquations2016,championUnifiedSparseOptimization2020} has provided practical ways to perform such model discovery. In contrast, few attempts exist for stochastic systems~\cite{boninsegnaSparseLearningStochastic2018, gaoAutonomousInferenceComplex2022,callahamNonlinearStochasticModelling2021,huangSparseInferenceActive2022,nabeelDiscoveringStochasticDynamical2024}, often lacking theoretical foundation and relying on heuristics. In this work, we establish a rigorous framework using extreme value statistics for model selection, providing a principled way to compare stochastic models and guide data-driven methods.

We organize this paper as follows. Our starting point is a quasi-likelihood maximization method, Stochastic Force Inference~\cite{frishmanLearningForceFields2020}, to estimate SDE parameters. We first show how likelihood estimates must be corrected in order to fairly compare two models, resulting in Akaike's information criterion (AIC) for SDEs. We then move to model selection from a library of basis functions, and show that AIC systematically fails to select the minimal model against the many more complex models due to multiple hypothesis testing. Our central result is a modified information criterion, Parsimonious Stochastic Inference (PASTIS), that combines exact results from likelihood estimation statistics and extreme value theory to select sparse SDE models at a chosen significance level. Importantly, PASTIS model selection accounts not only for the complexity of a given model, but also for the complexity-dependent combinatorial number of possible models. Comparing this method to pre-existing ones, we demonstrate that it performs comparably well in the near-deterministic sector and is a significant improvement over the state of the art for strongly stochastic systems. We show that it straightforwardly extends to continuous fields modeled by stochastic partial differential equations. Finally, we demonstrate the robustness of the method to data imperfections (sampling rate and measurement error), as well as its applicability to models of experimental interest: the identification of interaction networks in multi-species ecosystem and of chemical reaction-diffusion pathways.

\textbf{Model class.} We focus here on Brownian dynamics, the most broadly used class of continuous stochastic dynamical model. Specifically, we consider a $d$-dimensional autonomous first-order stochastic differential equation, 
\begin{equation}
    \dv{\mathbf{x}_t}{t} = \mathbf{F}(\mathbf{x}_t) + \sqrt{2\mathbf{D}(\mathbf{x}_t)} \mathbf{\xi}(t)
    \label{eq:Langevin}
\end{equation}
where the force field $\mathbf{F(x)}$ (also called \emph{drift}) characterizes the deterministic part of the dynamics, the diffusion matrix $\mathbf{D(x)}$ is symmetric positive definite and characterizes the stochastic part, and $\mathbf{\xi}$ is a $d$-dimensional Gaussian white noise. Throughout, we interpret multiplicative noise in the It\^o sense. Here we focus on the force field, which is generally the most physically informative part of the dynamics.

\textbf{Inference by linear regression.} Our goal is thus to reconstruct, from an observed time series $\mathbf{X}=\{\mathbf{x}_t\}_{t=0,\Delta t..., \tau}$, an inferred force field  $\mathbf{\hat{F}(x)}$ that best approximates the true $\mathbf{F}(\mathbf{x})$. To this aim, we adopt a widely used method consisting in approximating the force as a linear combination of basis functions $\mathcal{B}=\{\mathbf{b_i}(\mathbf{x})\}_{i=1..n_\mathcal{B}}$ with coefficients $\hat{F}_i$, such that the inferred force field reads 
\begin{equation}
    \hat{\mathbf{F}}^{\mathcal{B}}(\mathbf{x}) = \sum_{i=1}^{n_\mathcal{B}} \hat{F}_i^{\mathcal{B}} \mathbf{b}_i(\mathbf{x}).
    \label{eq:basis}
\end{equation} 
The inference problem thus decomposes into two parts: 1) selecting the basis functions $\mathbf{b_i}(\mathbf{x})$, which is the main focus of this article, and 2) inferring the corresponding coefficient values, for which we follow an approach closely related to Stochastic Force Inference~\cite{frishmanLearningForceFields2020} (SFI).

\textbf{Inferring coefficient values.} We first briefly summarize the SFI method. Our starting point is the following approximate log-likelihood function $\mathcal{L}(\mathbf{X}|\bar{\mathbf{F}} )$ for the trajectory $\mathbf{X}$ in a test force field $\bar{\mathbf{F}}$:
\begin{align}
    {\!\mathcal{L}(\mathbf{X}| \bar{\mathbf{F}})}= -\frac{\tau}{4} \avg{\!\left(\frac{\Delta \mathbf{x}_t}{\Delta t} - \bar{\mathbf{F}}_t\right)\!\cdot\!\mathbf{\bar{D}}^{-1}\!\cdot\! \left(\frac{\Delta \mathbf{x}_t}{\Delta t} - \bar{\mathbf{F}}_t \right)\!}
    \label{eq:likelihood}
\end{align}
where $\Delta \mathbf{x}_t = \mathbf{x}_{t+\Delta t} - \mathbf{x}_t$, $\bar{\mathbf{F}}_t = \bar{\mathbf{F}}(\mathbf{x}_t)$, and $\avg{\cdot} = \frac{1}{\tau} \sum_t \cdot \ \Delta t$ denotes trajectory averaging, with $\Delta t$ the time interval and $\tau$ the total time. Here $\bar{\mathbf{D}}= \frac{1}{2\Delta t} \avg{\Delta \mathbf{x_t} \otimes \Delta \mathbf{x_t}}$ is an estimate of the mean diffusion matrix. Importantly, when the dynamical noise is additive and with ideal data ($\Delta t \to 0$, no measurement error), we have $\mathbf{\bar{D}}\to \mathbf{D}$, and $\mathcal{L}$ coincides with the log-likelihood of the data in the force field $\bar{\mathbf{F}}$, up to an $\mathbf{\bar{F}}$-independent constant~\cite{riskenFokkerPlanckEquationSeveral1996}. In the general case of multiplicative dynamical noise, our approach remains practical, while true maximum likelihood is notoriously hard due to the difficulty of accurately estimating the state-dependent inverse diffusion matrix~\cite{siegertAnalysisDataSets1998,riskenFokkerPlanckEquationMethods1996,batzApproximateBayesLearning2018}.

In practice, given a basis of functions $\mathcal{B} = \{\mathbf{b_i}(\mathbf{x})\}_{i=1..n_\mathcal{B}}$, one can easily maximize \Eq{eq:likelihood} to obtain the inferred force coefficients 
\begin{equation}
\hat{F}_i^\mathcal{B} = \sum_j {G_\mathcal{B}^{-1}}_{ij} \avg{\frac{\Delta \mathbf{x}_t}{\Delta t}\cdot\mathbf{\bar{D}}^{-1}\cdot\mathbf{b_j}(\mathbf{x}_t)} 
    \label{eq:SFI-Ito}
\end{equation} where $(G_{\mathcal{B}})_{ij} = \avg{\mathbf{b}_i(\mathbf{x}_t)\cdot\mathbf{\bar{D}}^{-1}\cdot\mathbf{b}_j(\mathbf{x}_t)}$ is the Gram matrix associated with the basis $\mathcal{B}$. Note that in contrast with Ref.~\cite{frishmanLearningForceFields2020}, we define basis functions as vector functions and fitting coefficients are scalars, which provides a more flexible and general framework for basis selection. Once the coefficients $\hat{F}_i^\mathcal{B}$ are obtained, the force field can be reconstructed and extrapolated beyond the trajectory using \Eq{eq:basis}. In line with~\cite{frishmanLearningForceFields2020}, we can define the \emph{information} captured by the basis $\mathcal{B}$ as the log-likelihood gain of the inferred force field with the basis $\mathcal{B}$ compared to the null model with an empty basis (\emph{i.e.} pure Brownian motion with zero force),
\begin{equation}
    \mathcal{I}(\mathcal{B})  = \mathcal{L}(\mathbf{X}|\bo{\hat{F}}^\mathcal{B})- \mathcal{L}(\mathbf{X}|\bo{0}).
    \label{eq:I}
\end{equation}
This information, which can be evaluated from data only, serves as a starting point to estimate the quality of the fit with the basis $\mathcal{B}$.

\textbf{Comparing bases.} We now turn to the main question of this article: how to compare and select the basis functions? Comparing information $\mathcal{I}$ between bases is insufficient. Indeed, in this framework, we use the same data to estimate the force parameters $\hat{F}_i^{\mathcal{B}}$ and to evaluate the approximate likelihood $\mathcal{L}(\mathbf{X}|\hat{\mathbf{F}}^\mathcal{B})$, which biases selection towards overfitting: larger bases are favored, even if error increases. To overcome this, we aim to minimize the mean squared error between inferred and true force along the data, 
$\mathcal{E}(\hat{\mathbf{F}}^\mathcal{B}) = \frac{1}{4}\avg{\left(\mathbf{F} - \hat{\mathbf{F}}^\mathcal{B} \right)\cdot\bar{\mathbf{D}}^{-1}\cdot\left(\mathbf{F} - \hat{\mathbf{F}}^\mathcal{B} \right)}$. Although this error $\mathcal{E}$ is not accessible without knowing the true force $\mathbf{F}$, error differences between two bases $\mathcal{B}$ and $\mathcal{C}$ can be estimated: 
\begin{equation}
 \mathbb{E}[\mathcal{I}(\mathcal{C}) - \mathcal{I}(\mathcal{B})] \approx \tau\E{\mathcal{E}(\hat{\mathbf{F}}^{\mathcal{B}}) -  \mathcal{E}(\hat{\mathbf{F}}^{\mathcal{C}})} + n_{\mathcal{C}} - n_{\mathcal{B}}\label{eq:error}
\end{equation}
where $\mathbb{E}$ indicates expectation value over trajectory ensembles, $n_{\mathcal{B}}$ (resp. $n_{\mathcal{C}}$) is the number of functions in the basis $\mathcal{B}$ (resp. $\mathcal{C}$), and details are in \apdx~\ref{sec:E_from_log_likelihood}. The constant terms $n_{\mathcal{B}}$,$n_{\mathcal{C}}$ arise from noise correlations inducing a constant unit bias in $\mathcal{I}$ per parameter, characteristic of overfitting.

\begin{figure}[bt]
    \includegraphics[width=\linewidth]{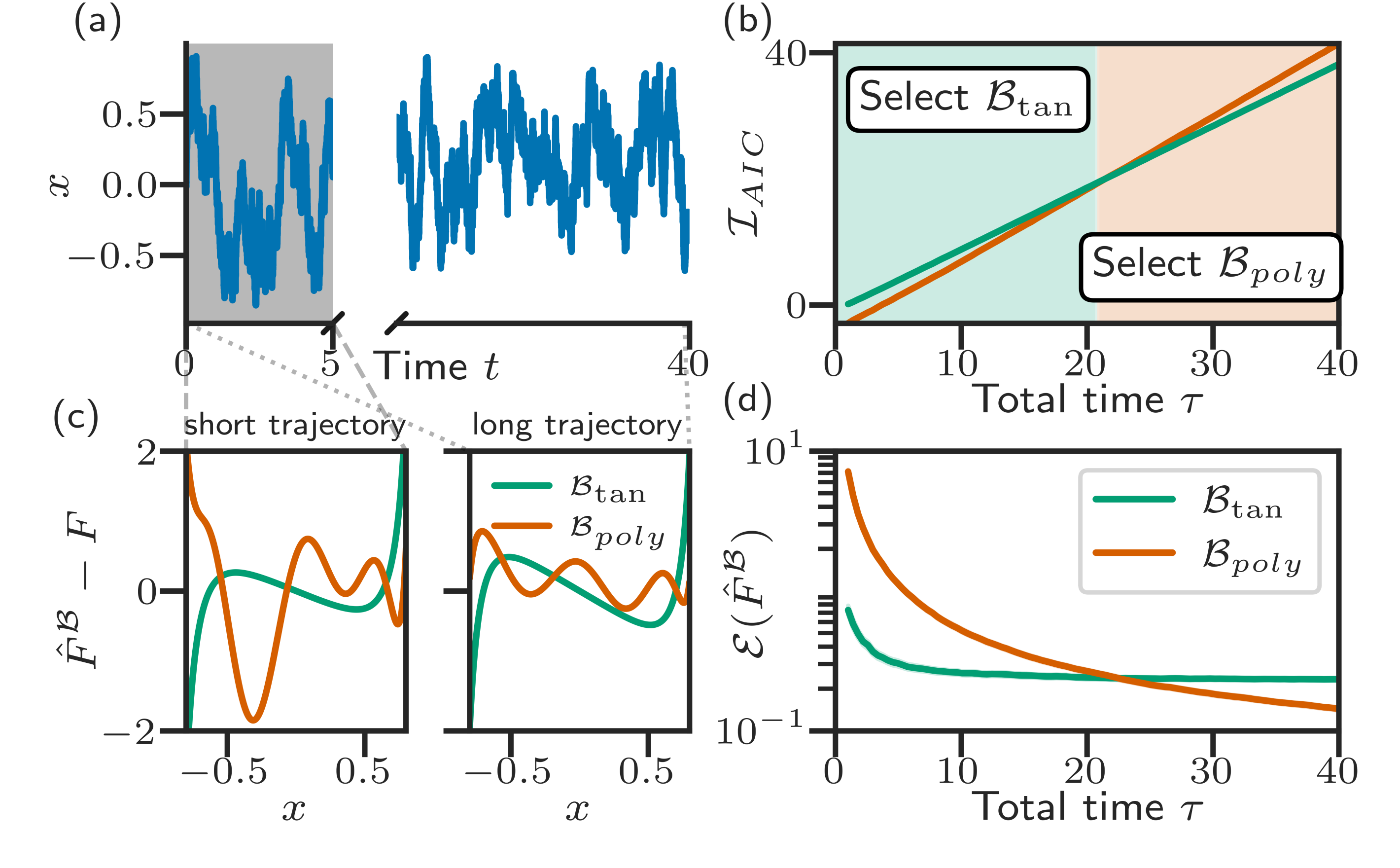}
    \caption{\textbf{Comparing models using AIC.} (a) Simulated trajectory of a one-dimensional toy model with force $F(x)=\frac{-x}{(1-x^2)^2}$ and dynamical noise $D=0.4$. (b)  $\mathcal{I}_\mathrm{AIC}$ averaged over $1000$ simulations as a function of total time for two bases: a single-parameter one $\mathcal{B}_{\tan}=\{\tan(x)\}$, and an order-8 polynomial $\mathcal{B}_{poly} = \{x^k\}_{k=0..8}$. (c) Inferred force minus true force, for each basis and for a short and a long trajectory. (d) Mean-squared error $\mathcal{E}$ between true and inferred force \emph{vs} total time.}
    \label{fig:AIC_exemple}
\end{figure}

\textbf{Akaike's information criterion.} To compare models, we can correct the bias in \Eq{eq:error} by defining $\mathcal{I}_{\mathrm{AIC}}(\mathcal{B}) = \mathcal{I}(\mathcal{B}) - n_\mathcal{B}$. This quantity coincides, up to a factor $-2$, with the Akaike Information criterion~\cite{akaikeNewLookStatistical1974}, a classic statistical estimator of model quality. On average, models with higher $\mathcal{I}_{\mathrm{AIC}}$ have a lower inference error along the trajectory: in particular, if $\mathcal{I}_{\mathrm{AIC}}(\mathcal{B})<0$, the model primarily fits the noise and a null model with zero force is better. As a practical example, in \Fig{fig:AIC_exemple}, we consider two possible bases: $\mathcal{B}_{\mathrm{tan}}$, with a single parameter that provides a simple but imperfect fit, and $\mathcal{B}_{\mathrm{poly}}$ with many parameters and which can provide a better fit. When the amount of data is low, we find that $\mathcal{I}_{\mathrm{AIC}}({\mathcal{B}_{\tan}}) >\mathcal{I}_{\mathrm{AIC}}({\mathcal{B}_{poly}})$ as the complex model overfits the data (\Fig{fig:AIC_exemple}b). In contrast, for large amounts of data, the complex model provides a better fit and $\mathcal{I}_{\mathrm{AIC}}({\mathcal{B}_{\tan}})< \mathcal{I}_{\mathrm{AIC}}({\mathcal{B}_{poly}}) $. Importantly, we confirm that the crossover between these two regimes coincides with the crossover in the actual inference error $\mathcal{E}$ (\Fig{fig:AIC_exemple}b,d). Thus, AIC model selection, which consists in choosing the model with the maximal $\mathcal{I}_{\mathrm{AIC}}$, estimated from data only, results in minimizing the true error $\mathcal{E}$.

\textbf{Sparse model selection.} When looking for a model without \emph{a priori} basis, a common technique is to start with a large yet finite library $\mathcal{B}_0$ of $n_0$ potential basis functions (e.g. polynomials, Fourier modes, etc). To avoid overfitting and enhance interpretability, such a library must then be reduced to a simpler basis $\mathcal{B}\subset\mathcal{B}_0$ by eliminating most of the functions. This reduction thus consists, in practice, in attempting to identify a simple, sparse model that best captures the data among the $2^{n_0}$ possible combinations of basis functions in the library. To do so, we propose to define an \emph{information criterion} allowing us to compare these many models, and to search among the possible models the one that maximizes this information criterion. Importantly, while AIC provides an unbiased way of comparing two models, it is not appropriate when comparing many models simultaneously.

\begin{figure}[t]
    \centering
    \includegraphics[width=\linewidth]{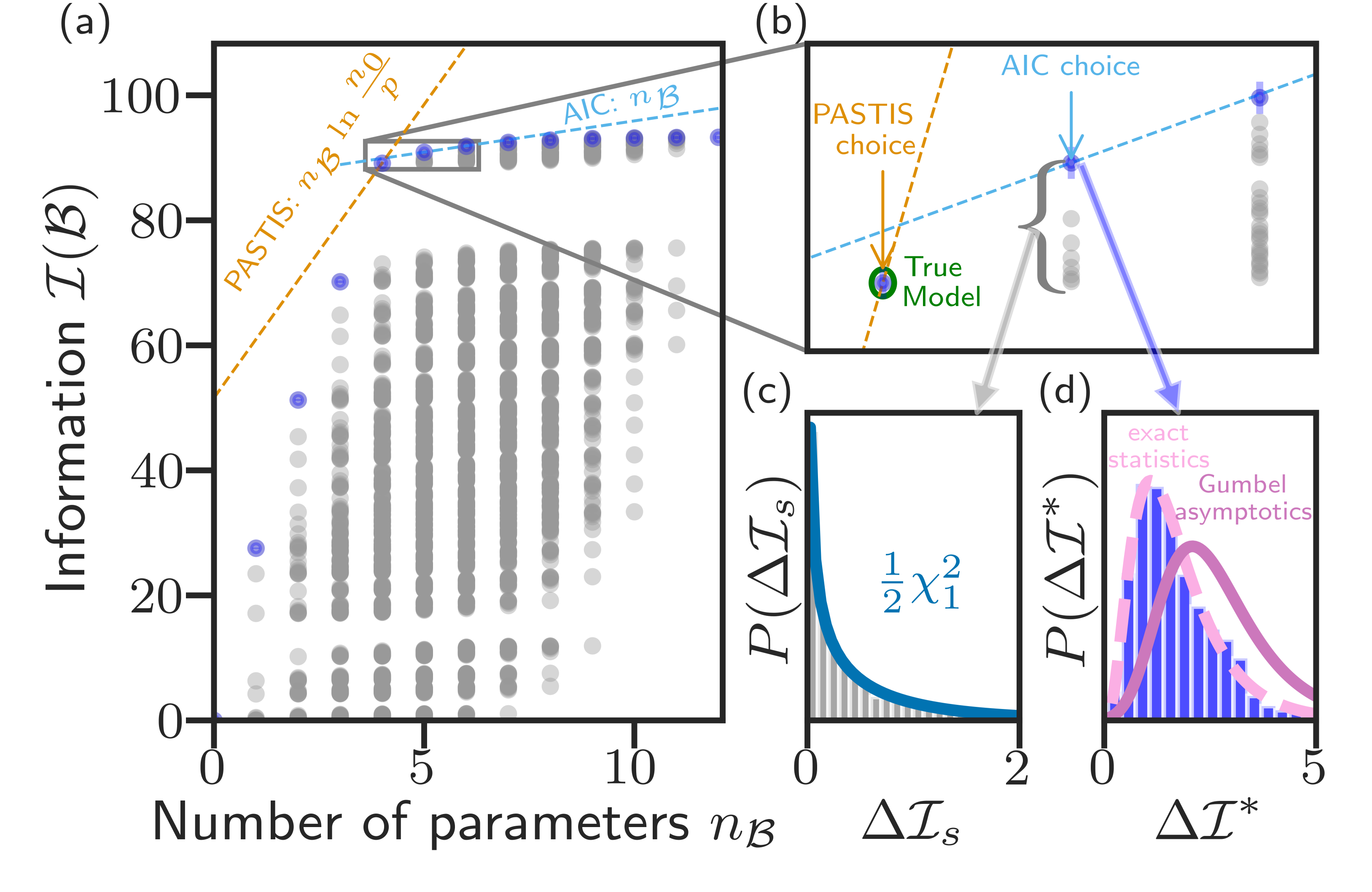}
    \caption{\textbf{Information statistics of sparse models.} (a) $\mathcal{I}(\mathcal{B})$ \emph{versus} $n_\mathcal{B}$ for all sub-models $\mathcal{B} \subset \mathcal{B}_0 = \{1 \mathbf{e_1}, .., x_3 \mathbf{e_3}\}$ (with $\mathbf{e_i}$ the unit vectors) of a 3-dimensional Ornstein-Uhlenbeck model with $n^*=4$ non-zero terms. The blue dots indicate the average $\mathcal{I}$ of the top-ranked model of size $n_\mathcal{B}$ across simulations, while gray dots correspond to the second, third, and lower-ranked models of size $n_\mathcal{B}$ across simulations. Dashed lines represent the slopes of the penalization terms for AIC and PASTIS, intersecting at the point corresponding to the selected model. (b) Close-up on (a), showing that AIC selects models with superfluous parameters, while PASTIS correctly identifies the true model $\mathcal{B}^*$. (c) The distribution of the information gains $\Delta\mathcal{I}_s$ for the true model + one superfluous term (histogram) is well captured by a $\chi^2_1$ distribution (solid line). (d) The maximum $\Delta \mathcal{I}^*$ of the information gain for one superfluous term $\Delta\mathcal{I}_s$ is well captured by analytical extreme values theory (dashed line, \apdx~\ref{sec:max_chi_square}) and converges to the Gumbel distribution (\Eq{eq:FTG}) when $n_0\to\infty$. All sub-figures are produced with $1000$ simulations.}
    \label{fig:Pastis_concept}
\end{figure}

\textbf{Failure of AIC.} Indeed, let us consider the well-specified case when there exists a true model $\mathcal{B}^* \subset \mathcal{B}_0$ containing $n^*$ functions that perfectly describe the system's dynamics. 
We find that with AIC, even in the limit of long trajectories, this true model is not recovered: more complex models are selected, with superfluous terms, as illustrated in \Fig{fig:Pastis_concept}a-b. This reflects a well-known limitation of AIC \cite{lebarbierIntroductionAuCritere2006}: it is not stringent enough to select the minimal or true model. To understand this, let us consider models consisting of the exact model plus one superfluous basis function, $\mathcal{B}^*+\{s\}$ for $s\in \mathcal{B}_0 - \mathcal{B}^*$. 
According to Wilks' theorem~\cite{wilksLargeSampleDistributionLikelihood1938}, the estimated log-likelihood difference asymptotically follows
\begin{equation}
\Delta \mathcal{I}_s = \mathcal{I}(\mathcal{B}^*+\{s\}) - \mathcal{I}(\mathcal{B}^*) \sim \frac{1}{2}\chi^2_1\label{eq:Wilks}
\end{equation}
as $\tau\to\infty$. This is indeed observed in practice (\Fig{fig:Pastis_concept}c). Thus $\E{\Delta \mathcal{I}_{\mathrm{AIC},s}}=\E{\Delta \mathcal{I}_{s}}-1= - \frac{1}{2}$: superfluous terms reduce the AIC value and tend to be eliminated on average. However, for a given superfluous term, there is a non-vanishing probability $P(\Delta \mathcal{I}_{\mathrm{AIC},s}> 0)\approx 0.157$ that the AIC difference is positive even in the limit of large data sets ($\tau\to\infty$). With many possible superfluous terms, the probability that one of them has an $\mathcal{I}_\mathrm{AIC}$ larger than that of $\mathcal{B}^*$ goes to one, hence the systematic failure of AIC to identify the true model (\Fig{fig:Pastis_concept}a-b).

\begin{figure*}[t]
     \centering
        \includegraphics[width=\textwidth]{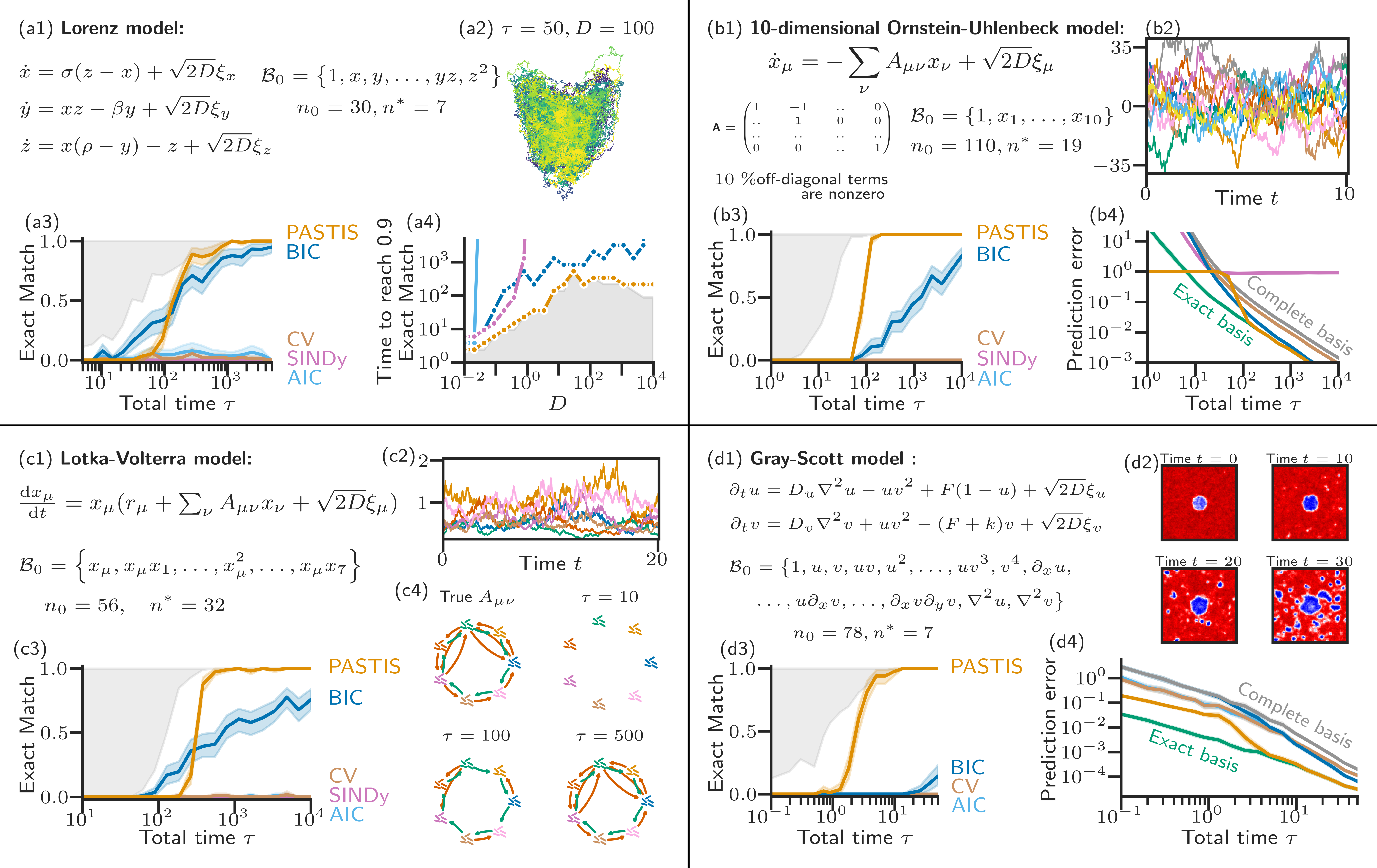}
        \caption{\textbf{Benchmarking PASTIS}. For the four models considered (a-d), we indicate (1) the generating equation and fitting basis $\mathcal{B}_0$ (each scalar function is considered along every unit vector),
        (2) a sample trajectory,  and (3) the exact match accuracy as a function of total time for different sparsity-enforcing algorithms. The gray area indicates cases when the true model does not maximize $\mathcal{I}$ between models with $n^*$ parameters. (a4): Trajectory time necessary to obtain $90\%$ exact match accuracy as a function of the Lorenz diffusion coefficient. (b4,d4): Prediction error of the inferred model for different algorithms, defined as $\mathcal{E}(\hat{\bo{F}}^\mathcal{B})/{\avg{\bo{F}\cdot(4\bo{\bar{D}})^{-1}\cdot\bo{F}}}$ computed on an independent, asymptotically long trajectory. (c4): True interaction network and reconstructed network as a function of time using PASTIS, for a sparse stochastic Lotka-Volterra model with environmental noise. All curves are averages over 48 simulations. Simulation details in \apdx~\ref{sec:simulations}.}
        \label{fig:benchmark}
\end{figure*}

\textbf{Extreme value statistics of the information.} Crucially, to identify the true model $\mathcal{B}^*$, we need to distinguish it from \emph{all} models with one superfluous term -- and in particular the one with the highest likelihood. We thus need to study the statistics of the information gap $\Delta\mathcal{I}^* = \max_{s\in\mathcal{B}_0-\mathcal{B}^*} \Delta \mathcal{I}_s$ between the true model and the $n_0 - n*$ models with one superfluous term. 
This extreme value problem can be tackled exactly (\apdx~\ref{sec:max_chi_square}) by assuming independence of the $\Delta \mathcal{I}_s$, \emph{i.e.} that $\avg{s(\mathbf{x}_t)s'(\mathbf{x}_t)}=0$ for superfluous functions $s\neq s'$. Under this assumption, the asymptotic behavior for large $n_0 - n^*$ is $\Delta\mathcal{I}^* \approx \log(n_0 - n^*) + Z$, where $Z\sim \mathrm{Gumbel}(\mu=0, \beta=1)$ is a standard Gumbel random variable. Using the properties of this distribution, we have 
\begin{equation}
    \mathbb{P}\left[\Delta\mathcal{I}^* < \log(n_0 - n^*) + z \right] \approx e^{-e^{-z}} \label{eq:FTG}
\end{equation} 
For finite $n_0-n^*$, this Gumbel approximation tends to overestimate $\Delta\mathcal{I}^*$ (\Fig{fig:Pastis_concept}d, \apdx~\ref{sec:demo_p_for_pastis}).

\textbf{Information criterion for large bases.} Based on these insights, we propose a modified information criterion, \emph{Parsimonious Stochastic Inference} (PASTIS), that includes the effect of extreme value statistics due to large libraries of functions:
\begin{equation}
\mathcal{I}_{\text{PASTIS}}(\mathcal{B}) = \mathcal{I}(\mathcal{B}) - n_\mathcal{B}  \log \frac{n_0}{p} 
\label{eq:PASTIS}
\end{equation}
Here, the user-chosen parameter $p \ll 1$ is a statistical significance threshold for accepting basis functions. It sets the target probability (approximately $p$) that the criterion would select a model containing one superfluous basis function rather than the true model (\apdx~\ref{sec:demo_p_for_pastis}). By selecting a small $p$, we demand strong evidence against the null hypothesis (that a term is superfluous) before including it, thus ensuring parsimony and controlling the risk of overfitting inherent in large-parametrized model selection. The multiple approximations made in the derivation of this criterion tend towards parsimony. In practice, we choose here $p=0.001$. Lowering this value further reduces the probability of overfitting, potentially at the cost of needing more data to identify all relevant coefficients in $\mathcal{B}_0$. The originality of $\mathcal{I}_{\text{PASTIS}}$ lies in explicitly accounting, in a principled way, for the size $n_0$ of the initial library in the complexity penalty. 

\textbf{Exploring Model Space.}
Maximizing $\mathcal{I}_{\text{PASTIS}}$ over the $2^{n_0}$ models is NP-hard. However, a greedy hill-climbing algorithm performs well: starting from an initial model, it accepts random single-parameter additions/removals if they increase $\mathcal{I}_\mathrm{PASTIS}$, until convergence. We run parallel searches initialized with null, full ($\mathcal{B}_0$), and random models. This rapidly finds the optimum when the true model maximizes $\mathcal{I}_\mathrm{PASTIS}$. The search is computationally efficient~\cite{supplement}.

\textbf{Benchmarking PASTIS.} We use synthetic data on four models to demonstrate PASTIS's efficiency: a stochastic Lorenz model (\Fig{fig:benchmark}a), a high-dimensional Ornstein-Uhlenbeck model with sparse coefficients  (\Fig{fig:benchmark}b), a generalized Lotka-Volterra model with multiplicative environmental noise and a sparse interaction network between species (\Fig{fig:benchmark}c), and a noisy Gray-Scott model for spatial reaction-diffusion dynamics (\Fig{fig:benchmark}d). In the first three cases, we use polynomial bases of first (b) and second (a,c) order. In \Fig{fig:benchmark}d, we consider a stochastic partial differential equation model, which we treat by enriching the basis with discretized differential operators, and consider all terms up to second-order derivatives and fourth order in the variables $u$ and $v$. To evaluate the performance of PASTIS and compare it with other approaches in identifying the correct model from these datasets, we use the \emph{exact match accuracy} defined as the fraction of independent simulation trials in which the algorithm selects a model $\mathcal{B}$ that is identical to the known true model $\mathcal{B}^*$. We find that, in all cases, with sufficient amounts of data, the exact match accuracy of PASTIS converges to a value $>1-p$ (\Figs{fig:benchmark}a3-d3). This criterion is near-optimal: in most cases where it fails to identify the true model, it is because another model with the same number of parameters has higher estimated likelihood (gray area in all panels of \Fig{fig:benchmark}), making the identification of $\mathcal{B}^*$ essentially impossible. 

\textbf{Comparing to other methods.} PASTIS compares favorably to alternatives  in terms of efficiently identifying exact models and minimizing prediction error (\Fig{fig:benchmark} and~\cite{supplement}). As discussed above, AIC and, for similar reasons, 7-fold Cross-Validation (CV), overfit and fail to identify exact models. Bayesian Information Criterion (BIC)~\cite{schwarzEstimatingDimensionModel1978,ahoModelSelectionEcologists2014,brollyBayesianComparisonStochastic2022} (adapted for SDEs, $\mathcal{I}_\mathrm{BIC}(\mathcal{B}) = \mathcal{I}(\mathcal{B}) - \frac{n_\mathcal{B}}{2} \log{\tau}$, see~\cite{supplement}) converges asymptotically, but much slower than PASTIS. Sparse Identification of Nonlinear Dynamics (SINDy)~\cite{bruntonDiscoveringGoverningEquations2016,kaptanogluPySINDyComprehensivePython2022} performs well for near-deterministic systems but degrades significantly with strong dynamical noise $D$ (\Fig{fig:benchmark}a4, STLSQ threshold=0.5). In misspecified situations where the exact force field cannot be expressed as a linear combination of functions from $\mathcal{B}_0$, PASTIS tends to select more parsimonious models than alternative methods, while retaining competitive prediction error~\cite{supplement}.

\begin{figure}[tb]
    \centering
    \includegraphics[width=\linewidth]{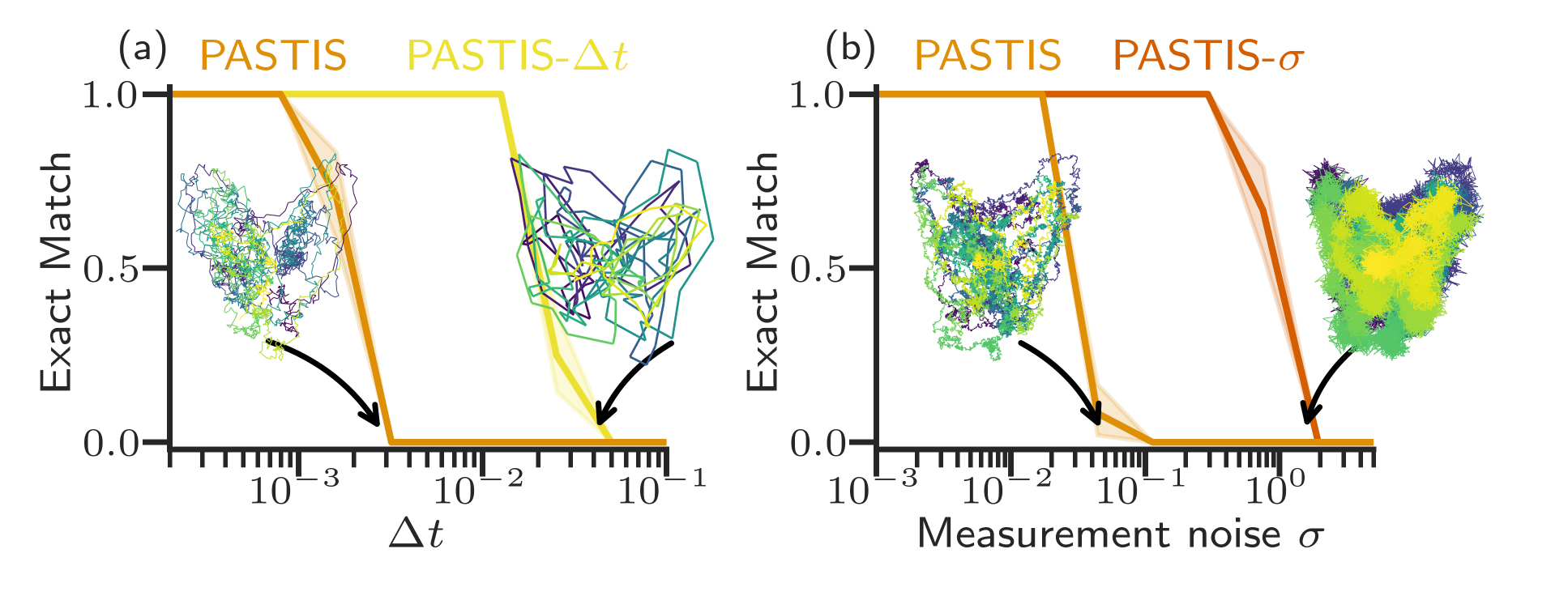}
    \caption{\textbf{Robustness of PASTIS.} (a) Exact match accuracy with and without trapezoid modification for large time intervals $\Delta t$, for the Lorenz model with long trajectories ($\tau=4\times10^4$). (b) Same 
 for the Stratonovich modification for measurement noise. Here $x_t \xrightarrow{} x_t + \eta$ where $\eta\sim\mathcal{N}(0,\sigma^2)$. Curves average over 48 simulations.}
    \label{fig:imperfect}
\end{figure}

\textbf{Data imperfection.} Experimental data typically have finite sampling intervals $\Delta t$ and measurement error. These imperfections incur biases both on the likelihood (\Eq{eq:likelihood}) and on the inferred force coefficients (\Eq{eq:SFI-Ito}). While a full treatment is complex, simple modifications significantly improve PASTIS: a trapezoid integration rule~\cite{amiriInferringGeometricalDynamics2024,wannerHigherOrderDrift2024} and a robust diffusion estimator for large $\Delta t$ (PASTIS-$\Delta t$ in \Fig{fig:imperfect}a), and a Stratonovich representation of stochastic integrals~\cite{frishmanLearningForceFields2020} plus robust diffusion estimation~\cite{vestergaardOptimalEstimationDiffusion2014} for measurement noise (PASTIS-$\sigma$ in \Fig{fig:imperfect}b). We give further details and explicit corresponding formulas in \apdx~\ref{appendix:data_imperfection}.

\textbf{Discussion.} We presented PASTIS, a principled method for sparse SDE model selection from data, based on quasi-likelihood maximization. Unlike methods relying on arbitrary thresholds~\cite{bruntonDiscoveringGoverningEquations2016,boninsegnaSparseLearningStochastic2018,madduStabilitySelectionEnables2022,kaptanogluPySINDyComprehensivePython2022}, penalization~\cite{tibshiraniRegressionShrinkageSelection1996,benmhenniGlobalOptimizationSparse2022}, or empirical significance testing~\cite{manganModelSelectionDynamical2017,frishmanLearningForceFields2020,brucknerInferringDynamicsUnderdamped2020,amiriInferringGeometricalDynamics2024}, PASTIS explicitly accounts for multiple hypothesis testing inherent in large function libraries. This perspective is consistent with extended Bayesian information criteria (EBIC), which add a combinatorial penalty to BIC for large model spaces~\cite{chen2008ebic}, and with Bayesian complexity priors that distribute prior mass over model sizes to adjust for multiplicity~\cite{scott2010multiplicity}. Leveraging exact results in likelihood estimation (Wilks' theorem, \Eq{eq:Wilks}) and extreme value statistics (\Eq{eq:FTG}), we derived the PASTIS criterion (\Eq{eq:PASTIS}), where the single parameter $p$ is an interpretable significance level. While we have kept this parameter fixed here, it could be made adaptive to both converge at long times and more efficiently fit at short times. We showed that this method is robust and efficient, even for high dimensions, dynamical noise, measurement error, and large time intervals. The inclusion of differential operators in the basis also permits the inference of stochastic partial differential equations from discretized fields, for which few inference methods pre-exist~\cite{mathpatiDiscoveringStochasticPartial2024}. This work provides a readily usable method~\cite{code} that paves the way towards direct inference of minimal models from experimental trajectories, for instance to identify biochemical pathways, ecological networks (\Fig{fig:benchmark}c) or reaction-diffusion mechanisms  (\Fig{fig:benchmark}d). This principled framework for sparse selection could be adapted for higher-order SDEs~\cite{brucknerInferringDynamicsUnderdamped2020,brucknerLearningDynamicsCell2021,ferrettiBuildingGeneralLangevin2020} and general likelihood-based inference problems~\cite{rishSparseModelingTheory2014}.

\begin{acknowledgments}
    We warmly thank Thierry Mora, Anna Frishman, Nicolas Levernier, Simon Gsell, Martin Lardy, Florence Bansept and Jo\~ao Valeriano for precious advice. 
    We acknowledge helpful mathematical input from ChatGPT. The project leading to this publication has received funding from France 2030, the French Government program managed by the French National Research Agency (ANR-16-CONV-0001) and from Excellence Initiative of Aix-Marseille University - A*MIDEX. PR thanks ICTP-SAIFR (FAPESP grant 2021/14335-0) where part of this work was done.
    Co-funded by the European Union (ERC-SuperStoc-101117322). 
\end{acknowledgments}

\bibliography{biblio,supplement}

\appendix
\section*{End Matter}

\subsection{Estimating the error $\mathcal{E}$ from the log-likelihood}
\label{sec:E_from_log_likelihood}
We prove here \Eq{eq:error} connecting the inference error $\mathcal{E}(\bo{\hat{F}}^{\mathcal{B}})$ to the estimated log-likelihood $\mathcal{L}(\bo{X}|\bo{\hat{F}}^{\mathcal{B}})$. For simplicity, we assume here that the normalization matrix $\bo{\bar{D}}$ is equal to the exact diffusion matrix $\bo{D}$, which has only a minor effect for non-multiplicative noise. We assume that $\Delta t$ is small enough to write $\Delta \bo{x}_t  \approx \bo{F}(\bo{x}_t)\Delta t + \Delta \bo{\Xi}_t$ where $\Delta \bo{\Xi_t} = \sqrt{2\bo{D}}\int_t^{t+\Delta t}\bm{\xi}(t) \dd{t}$. By expanding the log-likelihood, we find
\begin{multline*}
-\frac{4}{\tau}  \mathcal{L}\left(\bo{X} | \hat{\bo{F}}^\mathcal{B}\right) = \avgg{\left(\bo{F} - \hat{\bo{F}}^\mathcal{B}\right)} \\
+ \frac{2}{\Delta t}\avgs{\left(\bo{F} - \hat{\bo{F}}^\mathcal{B}\right)}{\Delta \bo{\Xi}_t} + \underbrace{\frac{1}{\Delta t^2}\avgg{\Delta \bo{\Xi}_t}}_{C}
\end{multline*}
where $C$ is model-independent and thus irrelevant for model comparison. We have $\E{\avgs{\bo{F}}{\Delta \bo{\Xi}_t}} = 0$. Applying the It\^o isometry:
\begin{multline*}
\label{eq:final_small_dt}
\E{\frac{2}{\Delta t} \avgs{\hat{\bo{F}}^\mathcal{B}}{\Delta \bo{\Xi}_t}} \approx\\ \E{\frac{4}{\tau}\sum_{i,j} \left({G_\mathcal{B}^{-1}}\right)_{ij}\avgs{ \bo{b_i} }{ \bo{b_j} }}
= 4\,\frac{n_{\mathcal{B}}}{\tau}.
\end{multline*}
where we neglected correlations between $\bo{G}_\mathcal{B}^{-1}$ and $\Delta \bo{\Xi}_t$ because they lead to higher-order terms. Consequently,
\begin{equation}
\label{eq:expected_logL}
\E{- \mathcal{L}\left(\bo{X} | \hat{\bo{F}}^\mathcal{B}\right)}= \tau \E{ 
    \mathcal{E}\left(\hat{\bo{F}}^\mathcal{B}\right) 
  } -  n_{\mathcal{B}} 
  + \E{C}
\end{equation}
which straightforwardly leads to \Eq{eq:error}. Using log-likelihood differences to estimate the error difference between models thus favors over-parameterized models.

\subsection{Statistics of the information gap $\Delta \mathcal{I}^*$}
\label{sec:max_chi_square}
Here, we study the distribution of $\Delta \mathcal{I}^*$ and prove its asymptotic Gumbel distribution (\Eq{eq:FTG}). We use the following result from (Ref.~\cite{leadbetterAsymptoticDistributionsExtremes1983}, example 1.7.4): for $N$ independent, identically distributed Gaussian random variables $X_1\dots X_N \sim \mathcal{N}(0,1)$, we have $P\left(\text{max}(X_1^2, \dots, X_N^2) \leq z\right) = \exp(-e^{-(z  - 2 \log(N))/2})$ to leading order when $N\to\infty$. Since each of the $n_0 - n^*$ variables $\Delta \mathcal{I}_s \sim \frac{1}{2} \chi_1^2$, and assuming their independence, we can apply the previous results to obtain an approximate cumulative distribution function of $\Delta\mathcal{I}^*$ (\Eq{eq:FTG}).  This parallels the “look-elsewhere effect” in large-scale searches, where the maximum over many null fluctuations also induces Gumbel-type corrections~\cite{gross2010trial,vitells2011estimating}.

Note that, with the hypothesis of independence between $\mathcal{I}_s$, we can go beyond this result to obtain the exact cumulative distribution function 
\begin{equation}
    \label{eq:exact_CDF}
    P(\Delta\mathcal{I}^*<z) = \erf{\left(\sqrt{z}\right)}^{n_0 - n^*}    
\end{equation} 
(pink dashed line in \Fig{fig:Pastis_concept}d), allowing for more refined estimation of the information gap.

\subsection{The parameter $p$ in $\mathcal{I}_{\text{PASTIS}}$}
\label{sec:demo_p_for_pastis}
\begin{figure}
    \centering
    \includegraphics[width=\linewidth]{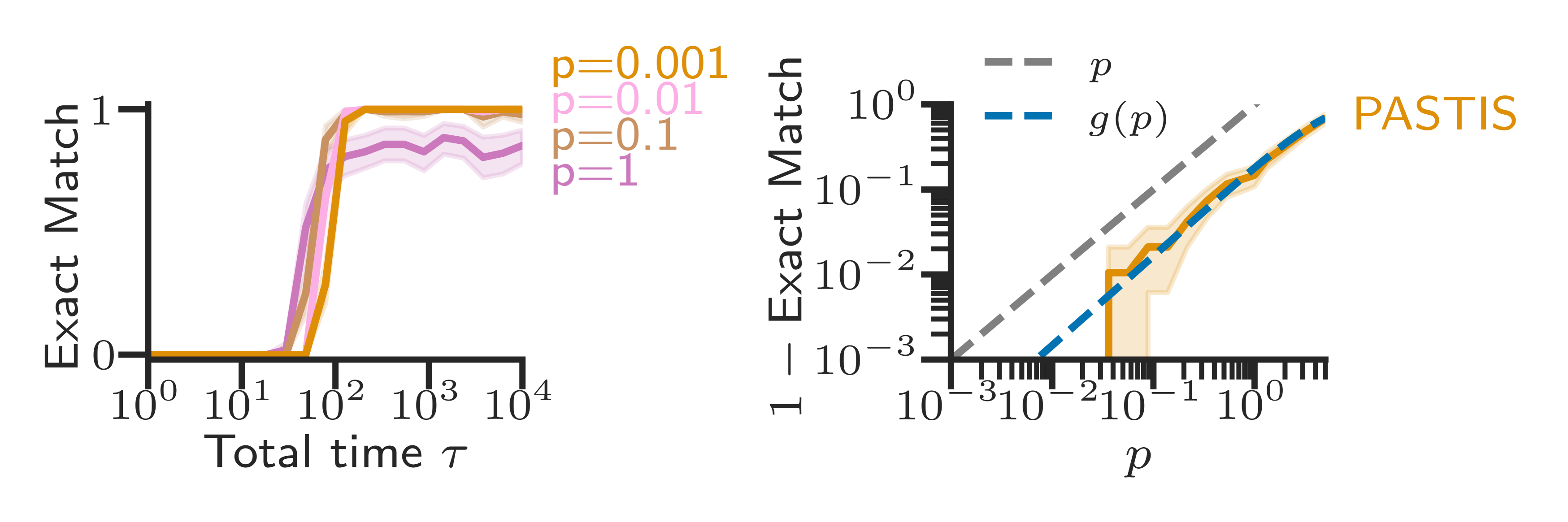}
    \caption{\textbf{Influence of $p$ on PASTIS} (a) Exact match accuracy for different value of $p$ for a 10-dimensional Ornstein-Uhlenbeck as in \Fig{fig:benchmark}b. (b) Asymptotic probability of identifying a wrong model against $p$ for a long trajectory with $\tau=10^4$. Blue dashed line: theoretical prediction $g(p) = 1 - [\erf{(\log{\frac{n_0}{p}})^{\frac{1}{2}}}]^{n_0 - n^*}$ for finite $n_0$. Curves are averaged over 100 simulations. More examples and details in~\cite{supplement}.}
    \label{fig:p_influence_OU}
\end{figure}

Here, we show that $p$, present in $\mathcal{I}_{\text{PASTIS}}(\mathcal{B})$ (\Eq{eq:PASTIS}), is the probability of selecting a model with one superfluous term which can be written as  $\mathbb{P}[\max_s \mathcal{I}_{\text{PASTIS}}(\mathcal{B}^*+\{s\}) > \mathcal{I}_{\text{PASTIS}}(\mathcal{B}^*)] \approx p$. First, we recall that $\max_s \mathcal{I}_{\text{PASTIS}}(\mathcal{B}^*+\{s\}) - \mathcal{I}_{\text{PASTIS}}(\mathcal{B}^*) = \Delta \mathcal{I}^* - \log\frac{n_0}{p}$. From \Eq{eq:FTG}, we obtain:
\begin{align*}
    \mathbb{P}[\Delta \mathcal{I}^* > \log{\frac{n_0}{p}}] &\approx 1 - \exp\left[-\frac{p (n_0 - n^*)}{n_0}\right]\\
    &\approx 1 - \exp[-p] \quad \text{with} \quad n_0 \gg n^*\\
    &\approx p \quad \text{with} \quad p \ll 1
\end{align*}
In the previous derivation, we replaced $n_0 - n^*$ by $n_0$, and we also used the approximated cumulative distribution function. These approximations tend to over-penalize complexity, as observed in \Fig{fig:p_influence_OU}b. A more precise penalization can be derived from the exact cumulative distribution function, leading to an accurate theoretical prediction of the exact match performance made by $\mathcal{I}_{\text{PASTIS}}$ for long observation time (curve $g(p)$ in \Fig{fig:p_influence_OU}b and~\cite{supplement}).
Increasing $p$ thus increases probability of exact model recovery in the long trajectory limit, at the cost of needing more data to start identifying the model (\Fig{fig:p_influence_OU}a). 

\subsection{Addressing Data Imperfections}
\label{appendix:data_imperfection}
Large sampling intervals $\Delta t$  and high measurement noise are two major challenges for both coefficient inference and model selection. We derive here modified estimators discussed in the main text and presented in \Fig{fig:imperfect}. Both rely on a Stratonovich transformation of the stochastic sum $\avg{\frac{\Delta \mathbf{x}_t}{\Delta t}\cdot\mathbf{\bar{D}}^{-1}\cdot\mathbf{b_j}(\mathbf{x}_t)}$ in the likelihood:
\begin{multline}
  \mathcal{L}_{St.}^{\hat{\bo{D}}}(\bo{X}|\bo{\bar{F}}) = -\sum_{\alpha,\beta,\gamma} \frac{\tau}{2} \avg{\hat{D}_{\gamma \beta }(\bo{x}_t)\pdv{\bar{F}_\alpha(\bo{x}_t)}{x_\beta}\langle\hat{D}\rangle^{-1}_{\gamma\alpha}}  -\\\tau \avg{\left(\left(\frac{\Delta \bo{x}_t}{\Delta t} - \frac{\bo{\bar{F}}(\bo{x}_{t+\Delta t}) + \bo{\bar{F}}(\bo{x}_{t})}{2}  \right)\frac{1}{\sqrt{4\langle\bo{\hat{D}}\rangle}}\right)^2}
   \label{eq:Likelihood-Strato}
\end{multline}
where we have explicit index summation in the first term, and $\bo{\hat{D}}(\bo{x}_t)$ is an instantaneous diffusion estimator. Note that when $\Delta t\to0$ and $\bo{\hat{D}}(\bo{x}_t)=\bo{{D}}(\bo{x}_t)$, \Eqs{eq:likelihood} and \ref{eq:Likelihood-Strato} are equivalent.
We now show how an adapted choice of $\hat{\bo{D}}$ improves the robustness of the method.

\paragraph*{Correcting large sampling intervals.}
When the sampling interval is large, we use a three-point estimator  $\bo{\hat{D}_{\Delta t}} = \frac{1}{4 \Delta t}(\Delta\bo{x}_{t} - \Delta \bo{x}_{t-\Delta t}) \otimes (\Delta\bo{x}_{t} - \Delta \bo{x}_{t-\Delta t})$ which removes the leading-order force-induced bias $\bo{F}^2\Delta t/2$. We complement this with a slight modification of the learned parameters:
\begin{equation}
        \hat{F}_{\Delta t,i}^\mathcal{B} = \sum_j \left({G_\mathcal{B}^{\Delta t}}\right)^{-1}_{ij} \avg{\frac{\Delta \mathbf{x}_t}{\Delta t}\cdot \langle\bo{\hat{D}_{\Delta t}}\rangle^{-1}\cdot \mathbf{b_j}(\mathbf{x}_t)} 
    \label{eq:SFI-Trape} 
\end{equation}
where $\left({G_\mathcal{B}^{\Delta t}}\right)_{ij} = \avg{\frac{(\bo{b_i}(\bo{x}_{t+\Delta t}) + \bo{b_i}(\bo{x}_t))}{2}\cdot \langle\bo{\hat{D}_{\Delta t}}\rangle^{-1} \cdot \bo{b_j}(\bo{x}_t)}$is a modified Gram matrix using trapezoid approximation, which has previously been shown to improve robustness of force estimation to large time intervals~\cite{amiriInferringGeometricalDynamics2024,wannerHigherOrderDrift2024}. We thus define the information criterion used in~\Fig{fig:imperfect}a as $\mathcal{I}_{\mathrm{PASTIS}-\Delta t} = \mathcal{L}_{St.}^{\bo{\hat{D}_{\Delta t}}}(\bo{X}|\bo{\hat{F}_{\Delta t}}^\mathcal{B}) - n_B \log{\frac{n_0}{p}}$.

\paragraph*{Correcting high measurement noise.}
We model measurement noise as $\bm{\eta}_t\sim\mathcal{N}(\bo{0}, \sigma\bo{I})$ that additively impacts the observed trajectory $\bo{x}_t \to \bo{x}_t + \bm{\eta}_t$. With classic estimators, this incurs $O(\sigma^2/\Delta t)$ biases, which are a major hindrance to force inference. This leading-order bias vanishes when using the Stratonovich formulation of the log-likelihood due to statistically telescoping terms~\cite{frishmanLearningForceFields2020} in the stochastic sum $\avg{\frac{\Delta \bo{\eta}_t}{\Delta t}\cdot \frac{\bo{F}^{\mathcal{B}}(\bo{x}_{t+\Delta t}) + \bo{F}^{\mathcal{B}}(\bo{x}_t)}{2}}$. We complement this with a corrected three-points estimator~\cite{vestergaardOptimalEstimationDiffusion2014} of the diffusion matrix $\bo{\hat{D}_{\sigma}} = \frac{1}{2 \Delta t} \Delta\bo{x}_{t} \otimes  \Delta\bo{x}_{t} + \frac{1}{\Delta t}\avg{\Delta \bo{x}_{t+\Delta t} \otimes \Delta \bo{x}_t}$. Then, by maximizing $\mathcal{L}_{St.}^{\bo{\hat{D}_{\sigma}}}$, we obtain the learned parameters:
\begin{multline}
    \hspace{-6mm}\hat{F}_{\sigma,i}^\mathcal{B} = \sum_j \left({G_\mathcal{B}^{\sigma}}\right)^{-1}_{ij} \bigg(\!\avg{\frac{\Delta \bo{x}_t}{\Delta t}\cdot\langle\bo{\hat{D}_{\sigma}}\rangle^{-1}\cdot\frac{\bo{b_j}(\bo{x}_t) + \bo{b_j}(\bo{x}_{t+\Delta t})}{2}} \\-  \sum_{\alpha,\beta,\gamma} \avg{(\hat{D}_{\sigma}(\bo{x}_t))_{\gamma \beta }{\pdv{b_{j,\alpha}(\bo{x}_t)}{x_\beta}}  \langle\hat{D}_{\sigma}\rangle^{-1}_{\gamma \alpha} }\!\bigg)
    \label{eq:SFI-Strato} 
\end{multline}
with $\left({G_\mathcal{B}^{\sigma}}\right)_{ij} = \avg{\bo{b_i}(\bo{x}_t)\cdot \bo{\langle\hat{D}_\sigma\rangle}^{-1} \cdot \frac{\bo{b_j}(\bo{x}_{t+\Delta t}) + \bo{b_j}(\bo{x}_t)}{2}}$. This estimator is closely related to the one previously introduced in Ref.~\cite{frishmanLearningForceFields2020}. We finally define the information criterion $\mathcal{I}_{\mathrm{PASTIS}-\sigma} = \mathcal{L}_{St.}^{\bo{\hat{D}_{\sigma}}}(\bo{X}|\bo{\hat{F}_{\sigma}}^\mathcal{B}) - n_B \log{\frac{n_0}{p}}$. Its minimization leads to the result presented in ~\Fig{fig:imperfect}b.

Note that correcting simultaneously for large $\Delta t$ and measurement noise is a substantial challenge for diffusion estimation, and thus for likelihood estimation and model inference.

\subsection{Sparse inference of stochastic partial differential equations}
We discuss here the adaptation of our information criterion $I_{\mathrm{PASTIS}}$ to Stochastic Partial Differential Equations (SPDEs) that we used in \Fig{fig:benchmark}d. We consider a two-dimensional field $\bm{\phi}(x,y,t)$ that follows:
\begin{equation}
    \pdv{\bm{\phi}(x,y,t)}{t} = F[\bm{\phi}] + \sqrt{2 D} \bm{\xi}(t, x, y)
\end{equation}
where $F[\bm{\phi}]$ is the force functional, and for simplicity we take $\xi$ to be an additive Gaussian white noise with $\E{\xi_\alpha(t, x, y)\xi_\beta(t, x, y)} = \delta(t-t')\delta(x-x')\delta(y-y')\delta_{\alpha\beta}$. We consider a discretized trajectory in space and time $\bm{\Phi} = \{\bm{\phi}(t_i, x_j, y_k)\}_{(t_i = 0, \cdots, \tau),(x_j=0,\cdots,L_x), (y_k=0,\cdots,L_y)}$. We adapt our trajectory average notation to fields with $\langle .\rangle = \frac{1}{\tau L_x L_y}\sum_{t_i,x_j,y_k}\cdot \Delta t \Delta x \Delta y$ where $L_x, L_y$ are spatial dimensions of the observed system. The log-likelihood is written for a test force functional $\bar{\bo{F}}$ : 
\begin{align}
   \!\!\!\! \!\mathcal{L}_\text{SPDE}( \bm{\Phi} | \bo{\bar{F}}) = \frac{-\tau L_x L_y}{4 \bar{D}} \avg{\!\!\left(\frac{\Delta \bm{\phi}}{\Delta t} - \bo{\bar{F}}\right)\!\cdot\!\left(\frac{\Delta \bm{\phi}}{\Delta t} - \bo{\bar{F}} \right)\!\!}
    \label{eq:likelihood_field}
\end{align}
where $\bar{D} = \avg{\frac{\left(\bm{\phi}(t+\Delta t, x, y) - \bm{\phi}(t, x, y)\right)^2}{2\Delta t}}$. We approximate the true force field $F$ using a linear combination of functionals $\mathcal{B} = \{b_i[\phi]\}_{i=1\dots n_{\mathcal{B}}}$, such as polynomials and differential operators of $\phi$. We discretize these operators using simple finite differences: $\pdv{\bm{\phi}}{x} = \frac{\bm{\phi}(t_i, x_j + \Delta x, y_k)- \bm{\phi}(t_i, x_j, y_k)}{\Delta x}$ and $\pdv{^2\bm{\phi}}{x^2} = \frac{\bm{\phi}(t_i, x_j + \Delta x, y_k)- 2\bm{\phi}(t_i, x_j, y_k) + \bm{\phi}(t_i, x_j - \Delta x, y_k)}{\Delta x^2}$. From the definition of the log-likelihood for SPDE (\Eq{eq:likelihood_field}), the logic developed in the main text is transferable from SDE to SPDE. Thus, we define our information criterion $I_{\mathrm{PASTIS}}$ for SPDE as: $I_{\mathrm{PASTIS}}(\mathcal{B}) = \mathcal{L}_\text{SPDE}(\bm{\Phi}|\bo{\hat{F}}^\mathcal{B}) - \mathcal{L}_\text{SPDE}(\bm{\Phi}|0) - n_\mathcal{B}  \log \frac{n_0 }{p}$ (\Fig{fig:benchmark}d). Note that we have not studied the $\Delta x \to 0$ limit here.

\subsection{Simulations details and parameters}\label{sec:simulations}
For all SDE and SPDE simulations, we use the Euler–Maruyama method with simulation time interval $\dd{t}$ and sampling time interval $\Delta t$, and total simulation time $\tau$. For the prediction error in \Fig{fig:benchmark},  we simulate a new independent trajectory with total time $\tau_{\mathcal{E}}$. Every trajectory is initiated using initial conditions obtained by simulating a trajectory for a total duration of $\tau_{therm.}=10$, starting from a random initial state. For Lorenz simulations (\Fig{fig:benchmark}a and \Fig{fig:imperfect}), we use $\sigma = 10,\rho=28, \beta=7/3$, $\dd{t}=\num{0.00001}$, $\Delta t = 0.0001$,  $D = 100$, $\tau_{\mathcal{E}}=20$ and only for \Fig{fig:imperfect}, $\tau=4 *10^3$. For each Ornstein-Uhlenbeck simulation, we choose a random $\bf{A}$ (\Fig{fig:benchmark}b1) with $1$ on the diagonal and 10\% of non-zero off-diagonal terms with $1$ or $-1$ (\Fig{fig:benchmark}b, \Fig{fig:p_influence_OU}). Then, we use $\dd{t}=0.001$, $\Delta t = 0.01$, $\tau = 10^4$, $\bo{D} = 100 \bo{I}$ where $\bo{I}$ is the identity matrix, $\tau_{\mathcal{E}}=10^3$. For the Lotka-Volterra model, we define the matrix $\bf{A}$ (\Fig{fig:benchmark}c1) with $-1$  on the diagonal and $1$ or $-1$ on the off-diagonal distributed as shown in \Fig{fig:benchmark}c4. For simulations, we use $\dd{t}=0.001$, $\Delta t = 0.01$, $\tau = 10^4$, $D = 0.05$, $\tau_{\mathcal{E}}=100$. For the Gray-Scott model, we use $D_u=0.2097, D_v=0.105, k=0.057 , F=0.029$. Then for simulations, we use a square lattice with periodic boundaries and discretized with $\dd{x}=\dd{y}= \Delta x =\Delta y = 1$, a length in space $L_x=L_y=100$, $\dd{t}=0.001$, $\Delta t = 0.01$, $\tau = 50$, $D = 0.001$, $\tau_{\mathcal{E}}=10$.

\clearpage\newpage

\onecolumngrid 

\begin{center}
\textbf{\large Supplemental Materials:\\ Principled model selection for stochastic dynamics}
\end{center}
\setcounter{equation}{0}
\setcounter{figure}{0}
\setcounter{table}{0}
\setcounter{page}{1}
\makeatletter
\renewcommand{\theequation}{S\arabic{equation}}
\renewcommand{\thefigure}{S\arabic{figure}}
\renewcommand{\bibnumfmt}[1]{[S#1]}
\renewcommand{\citenumfont}[1]{S#1}

\section{Choice and Interpretation of the PASTIS Parameter $p$}
\label{appendix:choice_p}

The PASTIS criterion $\mathcal{I}_{\text{PASTIS}}(\mathcal{B}) = \mathcal{I}(\mathcal{B}) - n_{\mathcal{B}} \log(n_0/p)$ (\Eq{eq:PASTIS}) has a single user-tunable parameter, $p$, which controls the stringency of the model selection penalty. As derived in Sec.~\ref{sec:demo_p_for_pastis} of the End Matter, $p$ approximately represents the target probability of incorrectly favoring a model containing at least one superfluous term over the true parsimonious model $\mathcal{B}^*$. This interpretation provides a principled way to choose $p$ based on the desired level of confidence against overfitting.

The choice of $p$ is analogous to selecting a significance level in standard hypothesis testing. A smaller $p$ corresponds to a stricter test (higher confidence required to include a term). For instance, choosing $p=0.01$ or $p=0.001$ reflects a strong preference for parsimony, similar to using high-$\sigma$ rules (e.g., $3\sigma$ corresponds roughly to $p \approx 0.003$) in conventional statistical tests. This contrasts with methods relying on arbitrary penalty tuning.

\begin{figure}[!htbp]
    \centering
    \includegraphics[width=0.9\linewidth]{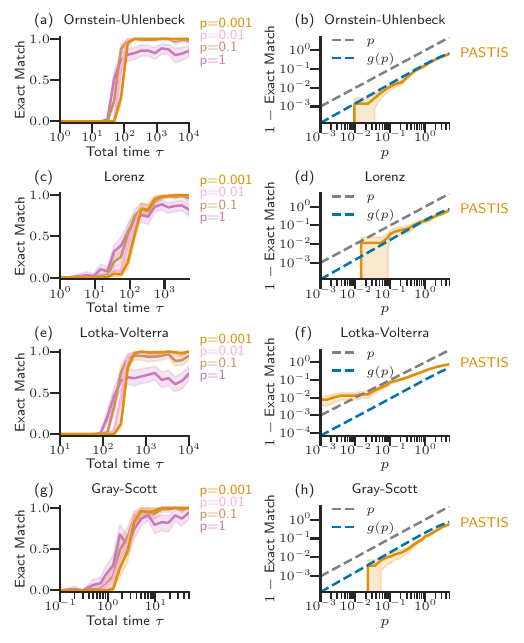} 
    \caption{Effect of the PASTIS parameter $p$ on the exact match accuracy for different benchmark systems. Left column (a, c, e, g): Exact match rate vs. trajectory time $\tau$ for fixed $p$ values (0.001, 0.01, 0.1, 1). Lower fixed $p$ requires more data but yields higher final accuracy. Right column (b, d, f, h): Empirical probability of error ($1 - \text{Exact Match}$, labeled "PASTIS") vs. $p$ (log scale), evaluated at a fixed large $\tau$. Dashed blue line: Theoretical prediction $g(p) = 1 - [\erf{(\log{\frac{n_0}{p}})^{\frac{1}{2}}}]^{n_0 - n^*}$ based on exact CDF for additive noise (see Sec.~\ref{sec:demo_p_for_pastis}). Curves are averaged over 100 simulations.}
    \label{fig:appendix_benchmark_p}
\end{figure}

The empirical impact of varying $p$ across the different benchmark systems is illustrated in \Fig{fig:appendix_benchmark_p}. The left column panels (a, c, e, g) show the exact match rate as a function of trajectory time $\tau$ for several fixed values of $p$. This clearly demonstrates the practical trade-off: lower values of $p$ (e.g., $p=0.001$, stricter penalty) require more data (longer $\tau$) to achieve high exact match rates, as the criterion is cautious about adding terms. Higher values of $p$ (e.g., $p=1$, weaker penalty) allow the model to achieve reasonable accuracy sooner but are more prone to overfitting, sometimes failing to reach perfect accuracy even with abundant data due to the persistent inclusion of superfluous terms.

The right column panels (\Fig{fig:appendix_benchmark_p}b, d, f, h) further explore the relationship between $p$ and the asymptotic probability of selecting a wrong model ($1 - \text{Exact Match}$, labeled "PASTIS") evaluated at a large $\tau$. The asymptotic probability is compared with the theoretical predictions $p$ and $g(p)= 1 - [\erf{(\log{\frac{n_0}{p}})^{\frac{1}{2}}}]^{n_0 - n^*}$ derived from the exact cumulative distribution function (see~\apdx~\ref{sec:demo_p_for_pastis}).

For systems dominated by additive noise (Ornstein-Uhlenbeck panel b, Lorenz panel d, Gray-Scott panel h), the theoretical prediction $g(p)$ reasonably matches the observed error rate. This comparison also highlights the nature of the approximation $g(p) \approx p$ used in the $I_{\text{PASTIS}}$ formula; while practical, the asymptotic probability of selecting a wrong model can differ slightly from $p$.

For the Lotka-Volterra system (\Fig{fig:appendix_benchmark_p}f), which features multiplicative noise, the theoretical prediction $g(p)$ (derived assuming additive noise) is less accurate, but the criterion remains pertinent and leads to high exact match rates when $p$ is small. 

\section{Benchmarking and Hyperparameter Tuning for Comparison Methods}
\label{appendix:benchmarks}

\begin{figure*}[!htbp] 
    \centering
    \includegraphics[width=\textwidth]{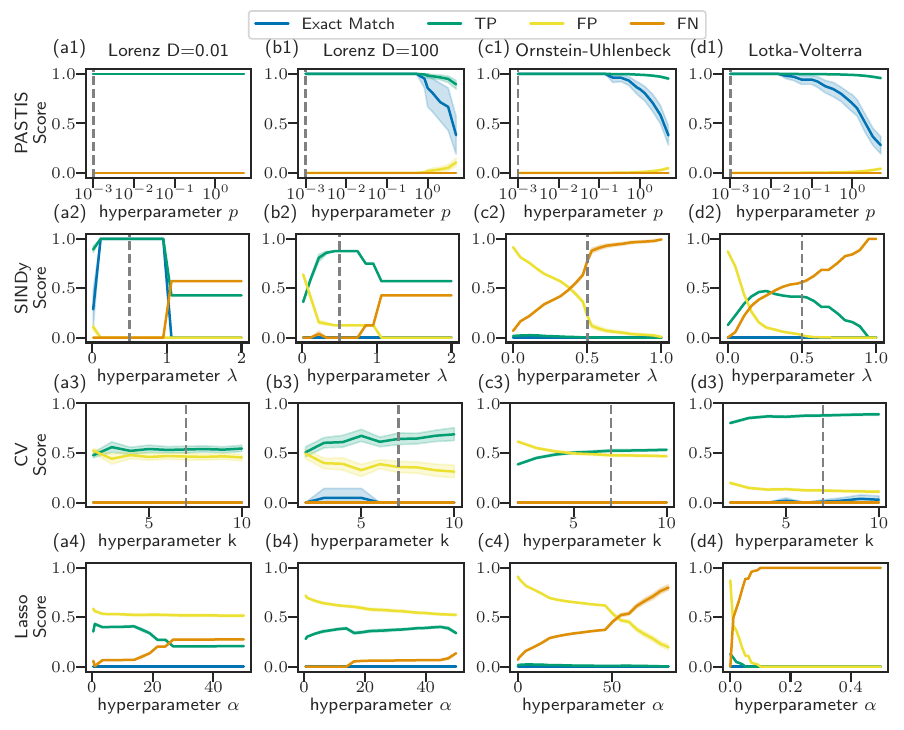}
    \caption{Comparison of model selection performance versus hyperparameters for PASTIS, SINDy, k-Fold CV, and LASSO across four benchmark systems (columns: Lorenz D=0.01, Lorenz D=100, Ornstein-Uhlenbeck, Lotka-Volterra). Each row corresponds to a method, plotting performance metrics against its primary hyperparameter: $p$ for PASTIS (row 1, panels a1-d1), STLSQ threshold $\lambda$ for SINDy (row 2, panels a2-d2), number of folds $k$ for CV (row 3, panels a3-d3), and regularization strength $\alpha$ for LASSO (row 4, panels a4-d4). Performance metrics are Exact Match (blue), True Positives (TP, green), False Positives (FP, yellow), and False Negatives (FN, orange). The vertical dashed lines indicate default or chosen parameter values used in main text comparisons where applicable (e.g., $p=0.001$ for PASTIS, $\lambda=0.5$ for SINDy). Note that the x-axis scale and range vary between rows according to the relevant hyperparameter. Curves are averaged over 20 simulations.}
    \label{fig:appendix_hyperparameters_total}
\end{figure*}

To ensure a fair comparison between PASTIS and existing methods (SINDy, k-fold CV, LASSO), we performed careful hyperparameter tuning for these benchmarks. \Fig{fig:appendix_hyperparameters_total} consolidates the results, showing performance metrics versus the primary hyperparameter for each method across the four benchmark systems. The metrics shown are Exact Match (blue), True Positives (TP, green), False Positives (FP, yellow), and False Negatives (FN, orange), defined as:
\begin{equation}
    \text{Exact Match} = \delta_{\mathcal{B},\mathcal{B}^*}, \qquad 
    \text{TP} = \frac{\lvert \mathcal{B} \cap \mathcal{B}^* \rvert}{\lvert \mathcal{B} \cup \mathcal{B}^* \rvert}, \qquad 
    \text{FP} = \frac{\lvert \mathcal{B} - \mathcal{B}^* \rvert}{\lvert \mathcal{B} \cup \mathcal{B}^* \rvert}, \qquad
    \text{FN} = \frac{\lvert \mathcal{B}^* - \mathcal{B} \rvert}{\lvert \mathcal{B} \cup \mathcal{B}^* \rvert},
    \label{eq:appendix_tp_fp_fn}
\end{equation}
where $\mathcal{B}$ is the model selected by the algorithm and $\mathcal{B}^*$ is the true model.

For the Akaike Information Criterion (AIC), Bayesian Information Criterion (BIC), and Cross-Validation (CV), the model space was explored using the same greedy forward-selection algorithm as employed for PASTIS. This iterative procedure is described in the main text. Algorithmic and implementation details for SINDy, CV, and LASSO are provided in their respective subsections below. Note that AIC and BIC are, in the formulas considered here, parameter-free.

\subsection{Bayesian Information Criterion}
\label{sec:BIC}
The Bayesian Information Criterion (BIC), against which our method is benchmarked in \Fig{fig:benchmark}, allows comparison of models in a Bayesian framework without having to select any particular prior $\Pi_\mathcal{B}(F^\mathcal{B}_1, \dots,F_{n_{\mathcal{B}}}^\mathcal{B})$ on the parameters $F^\mathcal{B}_i$. Indeed, it gives the asymptotic form of the marginal likelihood $P(\bo{X}|\mathcal{B})$ that is needed to obtain the posterior. We write the marginal likelihood for the model associated to the base $\mathcal{B}$:
\begin{equation}
    P(\bo{X}|\mathcal{B}) \!=\! \frac{1}{Z}\!\int\! e^{\mathcal{L}(\bo{X}|\bo{F^\mathcal{B}})} \Pi_\mathcal{B}(F^\mathcal{B}_1, \dots,F_{n_{\mathcal{B}}}^\mathcal{B}) \dd{F_1^\mathcal{B}}\dots\dd{F_{n_\mathcal{B}}^\mathcal{B}}
    \label{eq:posterior}
\end{equation}
where $\bo{F}^\mathcal{B} = \sum_i F_i^\mathcal{B} \bo{b_i}$, and $Z$ a $F_i^{\mathcal{B}}$-independent normalization constant. We Taylor-expand the likelihood (\Eq{eq:likelihood}) around the maximizing parameters $\hat{F}_i^\mathcal{B}$ of $\mathcal{L}$: 
\begin{equation}
\!\!\mathcal{L}(\bo{X}|\bo{F^\mathcal{B}}) \approx \mathcal{L}(\bo{X}|\bo{\hat{F}^\mathcal{B}}) - \frac{\tau}{8}(\bo{F}^\mathcal{B}-\hat{\bo{F}}^\mathcal{B})\cdot  \bo{G}_\mathcal{B} \cdot(\bo{F}^\mathcal{B}-\hat{\bo{F}}^\mathcal{B}) 
\end{equation}
By also expanding the prior $\Pi_\mathcal{B}$ around the maximizing parameters, injecting the previous result in \Eq{eq:posterior} and computing the integral, we obtain:
\begin{equation}
P(\bo{X}|\mathcal{B}) \approx \frac{e^{\mathcal{L}(\bo{X}|\bo{\hat{F}^\mathcal{B}})}}{Z} \left(\frac{8\pi}{\tau}\right)^{\frac{n_\mathcal{B}}{2}} (\det\bo{G}_\mathcal{B})^{-\frac{1}{2}}\Pi_\mathcal{B}(\hat{\bo{F}}^\mathcal{B})
\end{equation}
In the long trajectory limit $\tau \to \infty$, we have $(\det \bo{G}_\mathcal{B})^{-\frac{1}{2}}\Pi_\mathcal{B}(\hat{\bo{F}}^\mathcal{B}) = O(1)$. By taking the log of the marginal likelihood, neglecting this $O(1)$ term and model-independent constants, we find that the Bayesian Information Criterion ($BIC$) can be defined as :
\begin{equation}
\mathcal{I}_{\text{BIC}}(\mathcal{B}) = \mathcal{L}(\bo{X}|\bo{\hat{F}^\mathcal{B}}) - \frac{n_{\mathcal{B}}}{2}\log(\tau)
\end{equation}
Thus, comparing models by comparing BIC values results, asymptotically for $\tau \to \infty$, in the same conclusion as comparing the marginal likelihood for any prior. We note that in most textbooks, BIC is defined with the log of the number of data points instead of the log of total time, which is problematic when $\Delta t\to 0$. Our definition is consistent with the observation in Ref.~\cite{frishmanLearningForceFields2020} that the information per unit time is bounded in Brownian dynamics.

We note that, as a mitigation strategy for large libraries ($n_0$ big), the extended BIC (EBIC) augments the BIC penalty by an additional term $2\gamma \log \binom{n_0}{n_{\mathcal{B}}}$ (with $\gamma\!\in\![0,1]$ an empirical tuning parameter) to account for the number of candidate subsets and improve selection consistency as the model space grows~\cite{chen2008ebic}.

\subsection{SINDy (STLSQ)}
The SINDy algorithm used was based on the Python package \texttt{pysindy}, specifically commit \texttt{master@{2024-01-30 18:30:00}} \cite{kaptanogluPySINDyComprehensivePython2022}. We note that it is primarily designed for ordinary differential equation inference, in the absence of dynamical noise. We utilized the Sequential Thresholded Least Squares (STLSQ) optimizer with its default parameters, with the exception of the thresholding parameter $\lambda$, which was varied as shown in the second row of \Fig{fig:appendix_hyperparameters_total} (panels a2-d2). The optimal $\lambda$ varies with the system and noise level. For low noise (Lorenz D=0.01, panel a2), a sharp threshold exists where the exact match rate is 1. However, for noisier systems (panels b2-d2), the performance window disappears. Even near optimal $\lambda$ values (e.g., $\lambda \approx 0.5$ for Lorenz D=100, $\lambda \approx 0.1-0.5$ for OU and LV), SINDy often selects models with substantial numbers of both false positives (FP, yellow lines) and false negatives (FN, orange lines). This highlights its sensitivity and difficulty in achieving exact sparse recovery in challenging stochastic settings. The value $\lambda=0.5$ used in the main text represents a reasonable choice across the noisy systems tested.

\subsection{k-Fold Cross-Validation (CV)}
For CV, we employed a 7-fold cross-validation scheme. The trajectory was divided into 7 segments of equal length. For each fold, 6 segments were used as the training set to learn the model parameters (i.e., the coefficients $F_i^\mathcal{B}$ for a given basis $\mathcal{B}$), and the remaining segment was used as the validation set to compute the log-likelihood $\mathcal{L}(\bo{X}_{\text{valid}}|\bo{\hat{F}^\mathcal{B}}_{\text{train}})$. This process was iterated, with each of the 7 segments serving as the validation set once. The average log-likelihood across all folds was then used as an unbiased estimate of the model's predictive performance. The model $\mathcal{B}$ that maximized this average log-likelihood was selected.

The third row of \Fig{fig:appendix_hyperparameters_total} (panels a3-d3) shows the performance of k-fold CV versus the number of folds $k$ (where $k=7$ was used for the detailed evaluation just described, while the figure explores a range of $k$). The results are largely insensitive to $k$ in the range 2-10. More importantly, CV consistently fails to achieve a non null exact match rate for any $k$ across all systems. It primarily suffers from a high rate of false positives (FP, yellow lines), indicating that, like AIC, it does not sufficiently penalize model complexity when selecting from a large library in this SDE context.

\subsection{Least Absolute Shrinkage and Selection Operator (LASSO)}
For LASSO, we utilized the \texttt{pysindy} package (commit \texttt{master@{2024-01-30 18:30:00}}) as a framework for defining the feature library and problem structure. The LASSO regression itself was performed using the \texttt{Lasso} optimizer from the \texttt{sklearn.linear\_model} module in Scikit-learn. The fourth row of \Fig{fig:appendix_hyperparameters_total} (panels a4-d4) shows the performance of LASSO versus its regularization parameter $\alpha$. Similar to SINDy, performance is sensitive to $\alpha$. However, across the tested range, LASSO failed to achieve a non null exact match rate for any of the stochastic systems. It typically exhibits a trade-off where reducing false positives (by increasing $\alpha$) simultaneously increases false negatives, failing to identify the correct sparse structure. Its overall performance was significantly worse than PASTIS (row 1) and often worse than SINDy, and was not included in the main text for this reason. We note that similarly to SINDy, the poor performance of this method in stochastic settings is not necessarily surprising, as these methods are designed primarily for deterministic ODEs.

\subsection{Conclusion}

Based on the comprehensive hyperparameter analyses presented in \Fig{fig:appendix_hyperparameters_total}, the comparisons in the main text (\Fig{fig:benchmark}) are well-justified, using representative or near-optimal parameters for the benchmark methods. The figure clearly demonstrates the superior robustness and accuracy of PASTIS in achieving exact model recovery across different systems compared to SINDy, CV, and LASSO, especially when considering the challenge of minimizing both false positives and false negatives simultaneously.

\section{Performance in Misspecified Settings}
\label{appendix:misspecified}

In the benchmarks of PASTIS presented in the main text, we have focused on well-specified model identification scenarios, whereby a true underlying model  $\mathcal{B}^*$ exists and is a subset of the chosen library $\mathcal{B}_0$. This allowed us to focus on the exact match metrics. However, often the true underlying dynamics cannot be perfectly represented by any combination of functions within the chosen library $\mathcal{B}_0$. We investigate here the behavior of PASTIS in such scenarios.

A key strength of PASTIS is its penalty term $n_{\mathcal{B}} \log(n_0/p)$, which strongly discourages the inclusion of basis functions that do not significantly improve the likelihood. This property helps find the most parsimonious approximation within $\mathcal{B}_0$ that captures the essential dynamics without overfitting noise, even when the true model is outside the library.

\begin{figure}[!htbp] 
    \centering
    \includegraphics[width=0.49\linewidth]{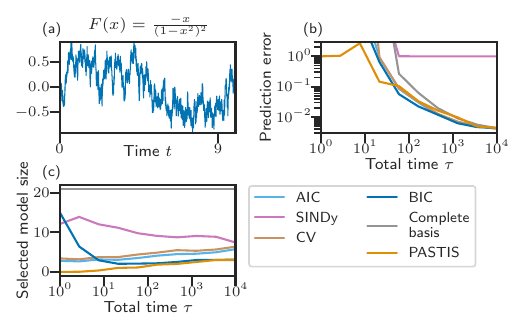} 
    \hfill 
        \includegraphics[width=0.49\linewidth]{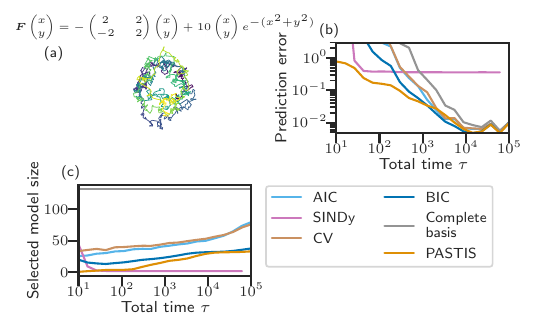} 
     \caption{Performance of PASTIS in misspecified model settings, using polynomial basis libraries of order 20 (left) and 10 (right) to infer non-polynomial dynamics. (a) A sample trajectory and the true force. (b) Prediction error on an independent trajectory of length $\tau_{\mathcal{E}} = 10^3$ (left) and $\tau_{\mathcal{E}} = 10^4$ (right) vs. total time $\tau$ for different model selection methods. (c) Selected model size vs. total time $tau$ for different methods. Curves average over 64 simulations.}
    \label{fig:appendix_misspecified_model}
\end{figure}

We considered two examples where the true force is non-polynomial, while the inference uses a standard polynomial basis library:
\begin{itemize}
    \item \textbf{Singular 1D Force} as in \Fig{fig:AIC_exemple} of the main text: A particle driven by the force $F(x) = -x / (1-x^2)^2$, approximated with a polynomial basis composed of monomials up to degree 20 (\Fig{fig:appendix_misspecified_model}, left). 
    \item \textbf{2D Ornstein-Uhlenbeck model with Gaussian Repulsion} as in Figs. 5-6 of Ref.~\cite{frishmanLearningForceFields2020}: A 2D system with linear dynamics plus a non-polynomial term (\Fig{fig:appendix_misspecified_model}, right), approximated with a polynomial basis composed of monomials up to degree 10.
\end{itemize}
\Fig{fig:appendix_misspecified_model} illustrates the results for these two cases, comparing PASTIS against other methods (AIC, BIC, CV, SINDy) and a model using the complete basis. Two metrics are used to assess the quality of the selected model: its prediction error (normalized mean squared error between true and inferred force along a long, independent simulated trajectory, panels (b)) and selected model size (panels (c)). In both cases, PASTIS selects smaller models than the other methods, while providing a competitive prediction error (only BIC performs slightly better in the intermediate total time regime). We conclude that while PASTIS is designed for well-specified settings, it also performs well and provides robust parsimonious models in misspecified settings.

\section{Computational Efficiency}
\label{appendix:efficiency}

Quite generally, PASTIS is computationally efficient, with each step of the model space exploration algorithm requiring only the inversion of the matrix $\mathbf{G}$ (\Eq{eq:SFI-Ito}). For instance, in the 10-dimensional Ornstein-Uhlenbeck model presented in \Fig{fig:benchmark}, the identification of a model with $n^* = 19$ terms in a basis with $n_0= 110$ functions can be reliably performed on a single CPU core (Intel® Xeon® E5-2670) in $\approx12$s.

\begin{figure}[!htbp]
    \centering
    \includegraphics[width=0.5\linewidth]{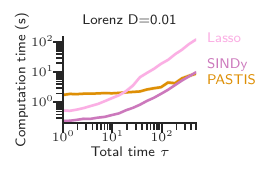} 
    \caption{Comparison of computation time (seconds, log scale) for PASTIS, SINDy, and Lasso applied to the Lorenz system benchmark ($D=0.01$) as a function of total trajectory length $\tau$. Curves are averaged over 20 simulations.}
    \label{fig:appendix_computation_time}
\end{figure}

We compare the computation time for PASTIS against SINDy and Lasso for the low-noise Lorenz system benchmark (D=0.01), running on Intel® Xeon® E5-2670. \Fig{fig:appendix_computation_time} shows the runtime as a function of trajectory length $\tau$.

As expected, the computation time for all methods increases with the trajectory length (number of data points). For this specific benchmark, PASTIS and SINDy exhibit very similar computational costs for large trajectory length. For small trajectory length, Lasso and SINDy appears notably faster, as PASTIS compute time becomes dominated by overhead from just-in-time compilation. However, the relevance of this comparison is limited since, as shown previously, Lasso and SINDy (in noisy cases) often fail to accurately identify the correct model structure.

\end{document}